\documentclass[prd,amsmath,aps,amssymb,nofootinbib,superscriptaddress,floatfix,twocolumn]{revtex4-2}
\usepackage{bm,amssymb,amsmath,siunitx,color,booktabs}
\usepackage[justification=raggedright,singlelinecheck=true]{caption}
\usepackage[mathscr]{euscript} 
\usepackage{graphicx}
\usepackage{upgreek}
\usepackage{dcolumn}
\usepackage{bm}
\usepackage[colorlinks=true,linkcolor=blue,urlcolor=blue,citecolor=blue]{hyperref}
\usepackage{float}
\usepackage{caption}
\usepackage{subcaption}
\usepackage{orcidlink}
\usepackage{makecell}
\usepackage{amsmath}
\begin{document}

\preprint{APS/123-QED}

\title{Calibration of key parameters during the in-orbit phase for the Taiji-2 gravitational reference sensor}

\author{Haoyue Zhang}
\thanks{These authors contributed equally to this work.}
\affiliation{Center for Gravitational Wave Experiment, Institute of Mechanics, Chinese Academy of Sciences, Beijing 100190, China}
\affiliation{State Key Laboratory of Micro-Spacecraft Rapid Design and Intelligent Cluster, Harbin Institute of Technology, Harbin 150001, China}%

\author{Chang Liu}
\thanks{These authors contributed equally to this work.}
\affiliation{Center for Gravitational Wave Experiment, Institute of Mechanics, Chinese Academy of Sciences, Beijing 100190, China}
\affiliation{Hangzhou Institute for Advanced Study, University of Chinese Academy of Sciences, Hangzhou 310024, China}
\affiliation{National Space Science Center, Chinese Academy of Sciences, Beijing 100190, China}%

\author{Xiaotong Wei}
\email{weixt@imech.ac.cn}
\affiliation{Center for Gravitational Wave Experiment, Institute of Mechanics, Chinese Academy of Sciences, Beijing 100190, China}%

\author{Peng Xu}%
\email{xupeng@imech.ac.cn}
\affiliation{Center for Gravitational Wave Experiment, Institute of Mechanics, Chinese Academy of Sciences, Beijing 100190, China}%
\affiliation{Hangzhou Institute for Advanced Study, University of Chinese Academy of Sciences, Hangzhou 310024, China}%
\affiliation{Lanzhou Center of Theoretical Physics, Lanzhou University, Lanzhou 730000, China}

\author{Li-E Qiang}%
\email{qianglie@nssc.ac.cn}
\affiliation{National Space Science Center, Chinese Academy of Sciences, Beijing 100190, China}%

\author{Ziren Luo}%
\affiliation{Center for Gravitational Wave Experiment, Institute of Mechanics, Chinese Academy of Sciences, Beijing 100190, China}%

\author{Ye Dong}%
\affiliation{State Key Laboratory of Micro-Spacecraft Rapid Design and Intelligent Cluster, Harbin Institute of Technology, Harbin 150001, China}%
\date{\today}

\begin{abstract}
The Taiji mission, a pioneering Chinese space-borne gravitational wave observatory, requires ultra-precise calibration of its gravitational reference sensors (GRSs) to achieve its targeted sensitivity of $3\times10^{-15} \mathrm{\ m\ s^{-2}\ Hz^{-1/2}}$. Maintaining this precision is challenged by time-varying scale factors drifts and dynamic center-of-mass (c.m.) offsets between the test masses (TMs) and spacecraft, driven by factors such as propellant consumption, thermal effects and aging electronics. This paper develops an advanced in-orbit calibration framework that simultaneously estimates the GRS scale factors and c.m. offsets between TMs and spacecraft through a combination of spacecraft maneuvers and Kalman filter. By applying periodic torque signals to induce controlled spacecraft angular accelerations, we leverage star tracker and GRS readouts to disentangle coupled disturbances and achieve dual-parameter calibration with unprecedented precision, with scale factors errors below 0.2\% and c.m. offsets residuals within 100 $\upmu$m, satisfies the Taiji-2 calibration requirements. This method is robust across different satellite configurations. The results not only ensure the feasibility of Taiji-2's scientific objectives but also establish a scalable calibration paradigm for future missions such as Taiji-3, where sub-micrometer c.m. stability and ultra-low noise gravitational reference will be essential. 
\end{abstract}
\maketitle

\section{Introduction}
\label{sec:intro}
In 2016, the Laser Interferometer Gravitational-Wave Observatory (LIGO) announced the direct detection of gravitational wave (GW) signals~\cite{PhysRevLett.116.061102}, ushering in the era of gravitational wave astronomy. 
Gravitational waves, which propagate through spacetime with minimal interaction with matter, offer
a unique probe of the universe, potentially covering the entire observable cosmos. However, ground-based observatories like LIGO are limited by terrestrial noise and the finite baseline of their interferometers, restricting their sensitivity to frequencies between 10 Hz and 1 kHz. To overcome these limitations, space-based GW observatories have been proposed and developed, such as the Laser Interferometer Space Antenna (LISA)~\cite{amaro2017laser}, Taiji~\cite{10.1093/nsr/nwx116} and Tianqin~\cite{luo2016tianqin}.

Taiji mission~\cite{10.1093/nsr/nwx116,10.1038/531150a}, led by the Chinese Academy of Sciences, is a Chinese space-based gravitational wave detection, aimed at detecting GW in the frequency range of 0.1 mHz to 1 Hz~\cite{10.1093/ptep/ptaa083}. 
This frequency range encompasses a wide variety of sources such as (super)massive black hole mergers, extreme mass ratio inspirals, compact binary systems, and stochastic gravitational wave backgrounds~\cite{doi:10.1142/S0217751X2050075X,Ruan2020,Barausse_2015,PhysRevD.78.124011,Postnov_2014,PhysRevLett.48.1220}. These sources are essential for enhancing our understanding of the universe's evolution and advancing astronomical research~\cite{doi:10.1142/S0217751X2050075X}.
Taiji mission will deploy three spacecraft (SC) in an equilateral triangle, each housing two free-falling test masses (TMs) as gravitational references, GW will be detected by measuring the relative displacement between test masses using space-based laser interferometry.
To meet the scientific objectives of the Taiji mission, the residual acceleration noise of each TM along the sensitive axis must remain below $3\times10^{-15}\ \mathrm{m}\ \mathrm{s}^{-2}\ \mathrm{Hz}^{-1/2}$.

Taiji-1 technology demonstration satellite achieved a gravitational reference sensor (GRS) measurement noise level of $2\times10^{-9}\ \mathrm{m}\ \mathrm{s}^{-2}\ \mathrm{Hz}^{-1/2}$~\cite{wang2021development}. Building on this, Taiji-2 mission aims to reduce the residual acceleration noise to $3\times10^{-14}\ \mathrm{m}\ \mathrm{s}^{-2}\ \mathrm{Hz}^{-1/2}$. Achieving this level of precision requires overcoming significant challenges, including the calibration of key parameters such as the GRS scale factors that convert the voltage readouts into physical linear and angular accelerations of TM along each degree of freedom and the center-of-mass (c.m.) offsets between the TM and the SC~\cite{zhang2023systematic,PhysRevD.108.082001}. These parameters are susceptible to time-varying drifts caused by factors such as propellant consumption, thermal effects, and aging of electronic components, which can degrade the performance of the GRS and compromise the mission's scientific objectives.
In this work we distinguish two operational modes of the GRS. In the accelerometer mode, the TM is electrostatically suspended around its nominal position. 
In the science mode, the TM undergoes geodesic motion along the sensitive axis and the spacecraft is driven by micro-thrusters to follow the TM.

This work presents the first in-orbit calibration scheme tailored to the Taiji-2 spacecraft configuration, together with a dedicated numerical simulation that verifies the achievable accuracies for both the GRS scale factors and the TM–SC c.m. offsets. 
In this paper, we present an advanced in-orbit calibration framework that simultaneously estimates the GRS scale factors and TM c.m. offsets through a combination of SC maneuvers and Kalman filter. By applying periodic torque signals to induce controlled angular accelerations, we utilize star tracker (ST) and GRS readouts to decouple cross-coupled simultaneous calibration of both the scale factors and c.m. offsets.
The method achieves high precision, with scale factors errors below 0.2\% and c.m. offsets residuals within 100 µm, and demonstrates robustness across varying spacecraft configurations. 
Our results not only ensure the feasibility of Taiji-2's scientific objectives but also establish a scalable calibration framework for future missions like Taiji-3, where sub-micrometer c.m. stability and ultralow residual acceleration gravitational reference will be essential. This work highlights the critical importance of dynamic in-orbit calibration in advancing space-based gravitational wave detection.

This paper presents the in-orbit scale factors and c.m. calibration experiment scheme and simulation results of Taiji-2.
Section~\ref{sec:dynamics} describes the GRS measurement principle and system dynamics, as well as the calibration accuracy requirements for the scale factors and c.m. offsets. 
Section~\ref{sec:offset_scheme} elucidates the in-orbit calibration methodology for  the scale factors and c.m. offsets. 
Section~\ref{sec:result} presents the in-orbit estimation results,  while Section~\ref{sec:discussion} provides conclusions and discussion.

\section{measuring principle and dynamics of Gravitational Reference Sensor}
\label{sec:dynamics}
\begin{figure}[htb]
\centering
\includegraphics[width=.4\textwidth]{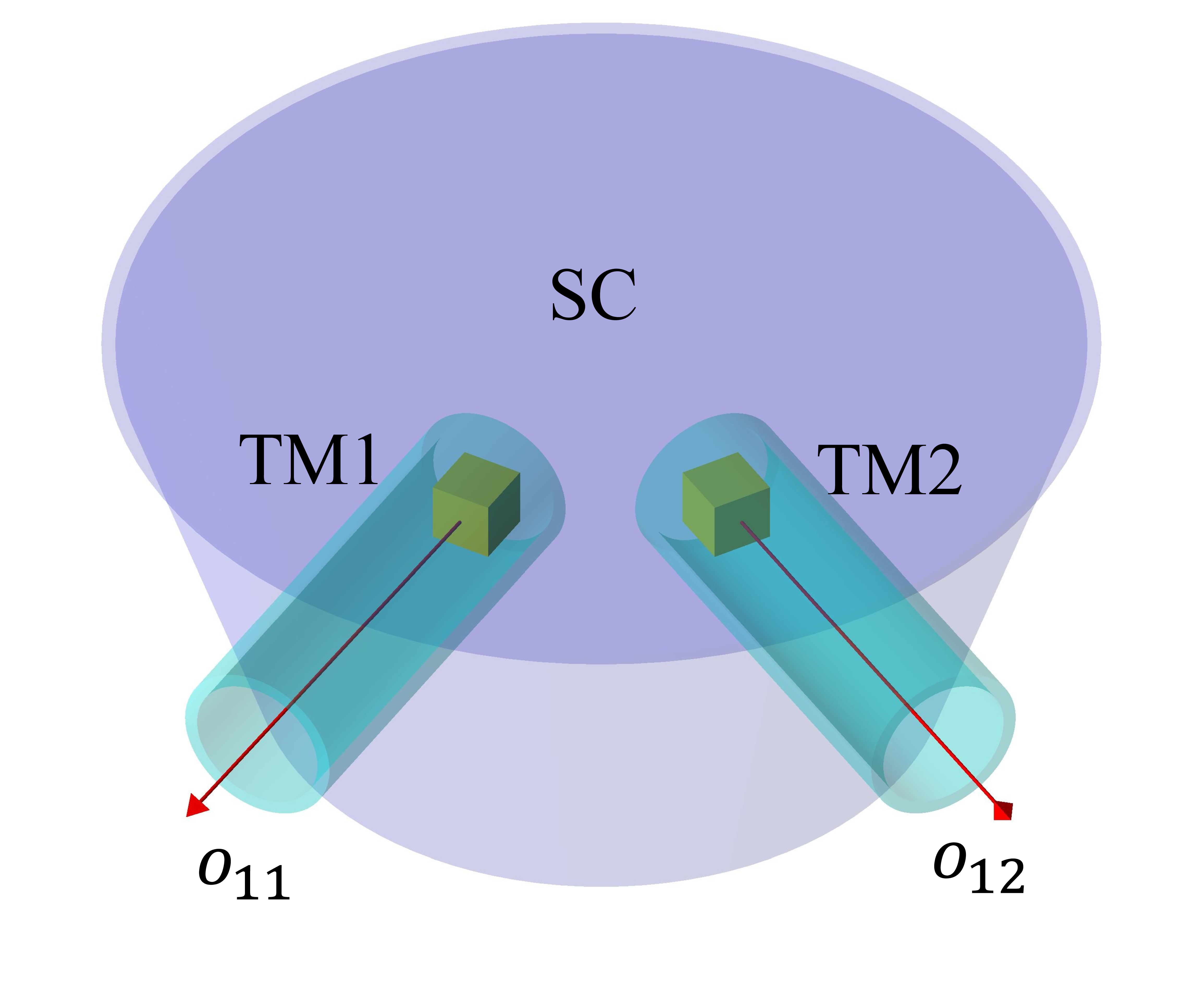}
\caption{The geometric structure of the Taiji-2.}
\label{fig:taiji2}
\end{figure}

In Taiji-2 mission, each SC is equipped with two GRSs, detailed in Fig.~\ref{fig:taiji2}.
The GRS uses capacitive sensing to measure non-gravitational acceleration disturbances acting on Taiji-2. 
The data collected are sent to the drag-free control system, which commands the micro-thrusters (as the drag-free actuators) to compensate for non-gravitational acceleration disturbances along the sensitive axis~\cite{doi:10.1142/S0217751X21400194}. 
Consequently, the accuracy of the GRS readouts is vital for the drag-free control system to ensure that the SC follows the motion of the TMs along the sensitive axis.

\begin{figure}[htb]
\centering
\includegraphics[width=.4\textwidth]{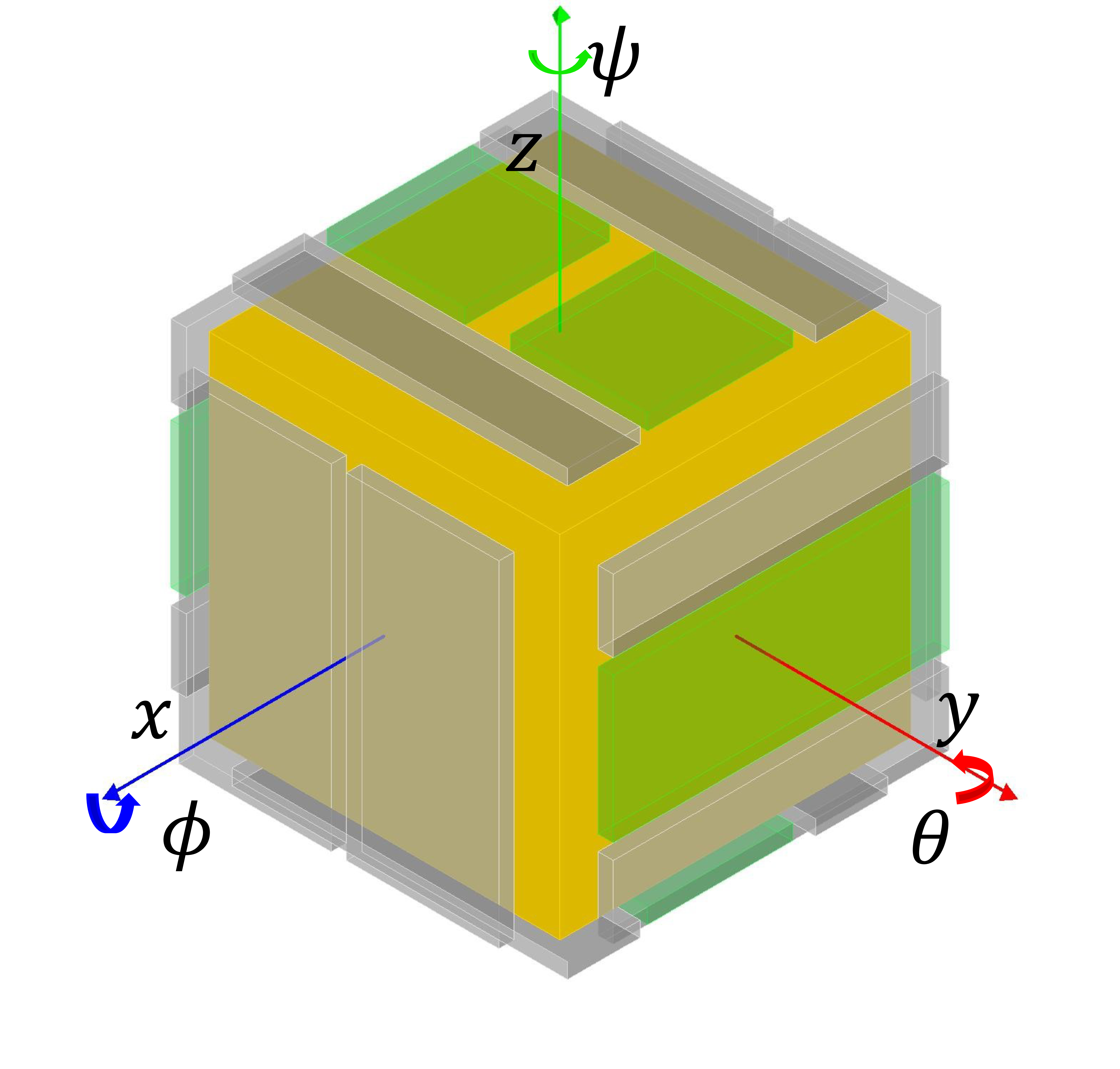}
\caption{Layout of the electrodes of the GRS.}
\label{fig:Taiji_XYZ}
\end{figure}

\begin{figure}[htb]
\centering
\includegraphics[width=.4\textwidth]{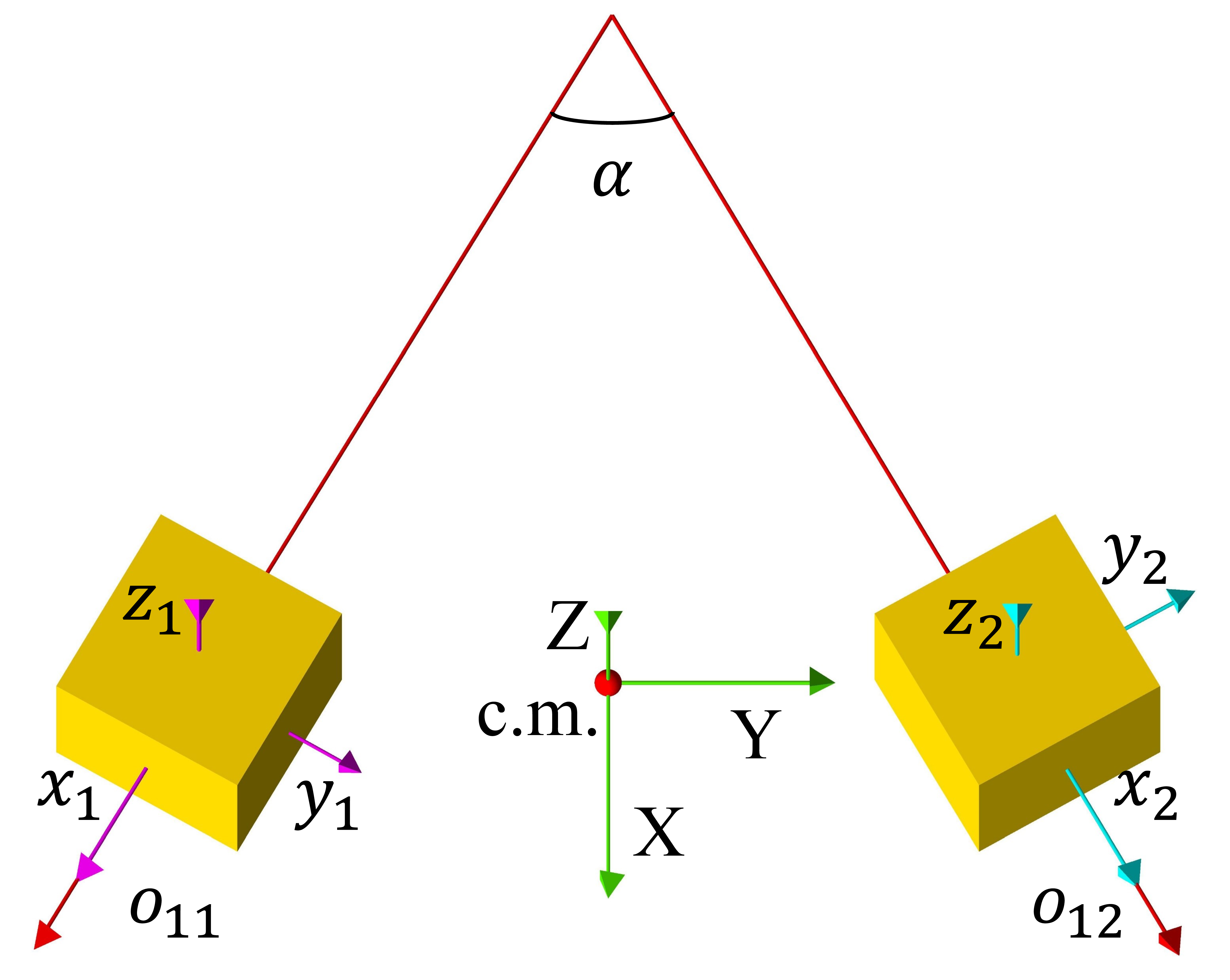}
\caption{SC c.m. location and TM distribution (Type-1).}
\label{fig:Pos_com1}
\end{figure}

\begin{figure}[htb]
\centering
\includegraphics[width=.4\textwidth]{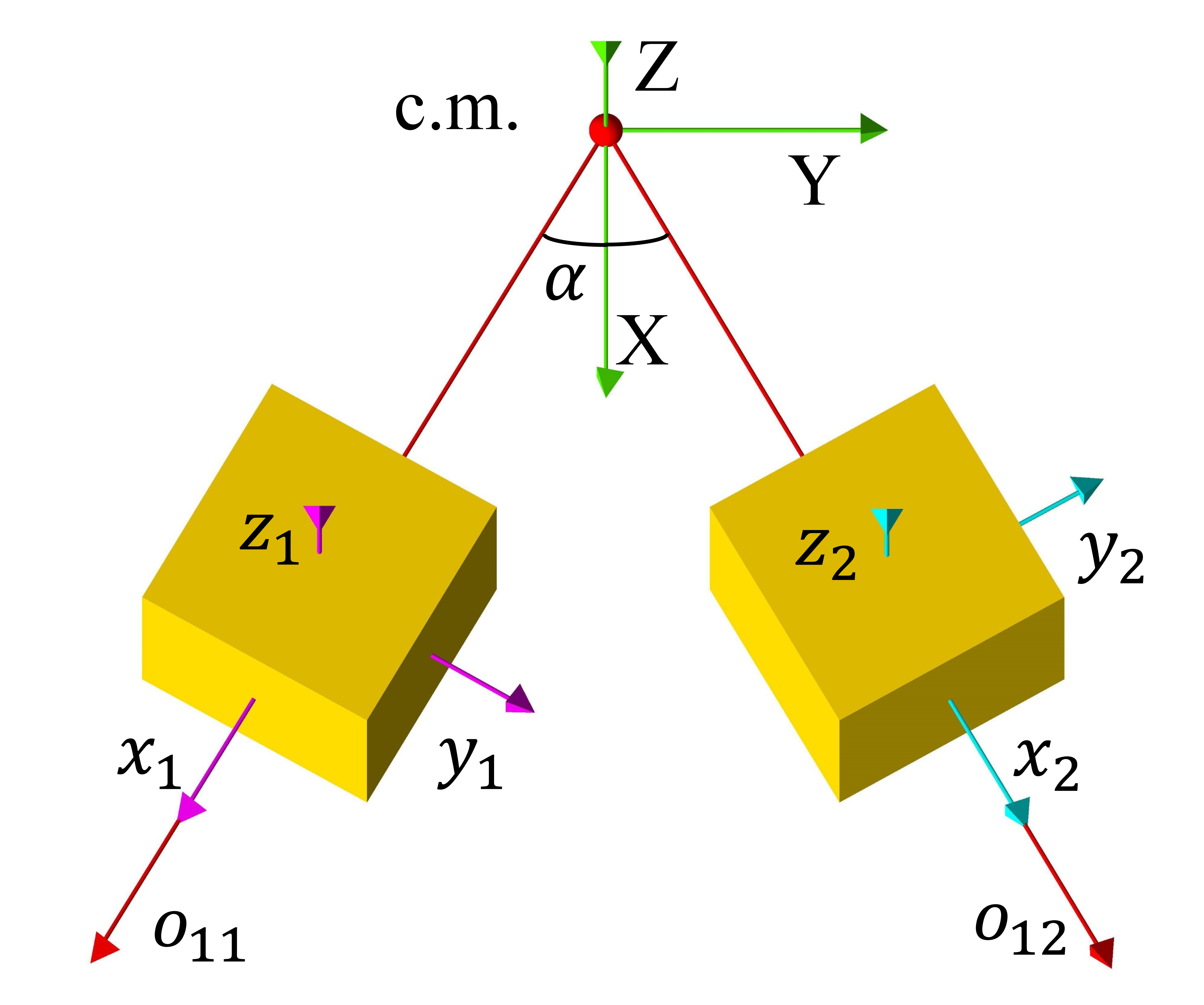}
\caption{SC c.m. location and TM distribution (Type-2).}
\label{fig:pos_SC_com1}
\end{figure}

The electrostatic system of the GRS is housed within a vacuum chamber and consists of a mechanical assembly and a front-end electronics (FEE) unit. 
The mechanical assembly includes a 1.96 kg titanium-alloy cubic TM measuring $46\ \mathrm{mm} \times 46\ \mathrm{mm} \times 46\ \mathrm{mm}$, encased in an electrode cage.
Both the TM and its enclosure cage are coated with gold. 
Within the cage, six pairs of electrodes are positioned opposite the surfaces of the TM, a schematic illustration and the definition of the measurement axes within the GRS are provided in Fig.~\ref{fig:Taiji_XYZ}. 
During in-orbit operation, changes in the position of the TM relative to the cage lead to variations in the capacitance of electrodes. These variations generate signals via the sensing bridges~\cite{wang2021development}, which are detected by the FEE and used to determine the position and attitude of the TM. Based on this information, the TM is actively maintained at its designated position through servo control.
The GRS nominal precision level of Taiji-2 is $3\times10^{-14}\ \mathrm{m}\ \mathrm{s}^{-2}\ \mathrm{Hz}^{-1/2}$, Tab.~\ref{tab:IS} gives the key design requirements. 
\begin{table}[b]
\caption{Key design requirements of Taiji-2 GRS. Scientific mode range refers to the measurement range of the GRS in science mode.}
\label{tab:IS}
\begin{ruledtabular}
\begin{tabular}{cc}
Parameter &   Nominal Value\\
\hline
Scientific mode range    &    $(-1\sim 1)\times 10^{-9}\ \mathrm{m}\ \mathrm{s}^{-2}$\\
Bandwidth      &      $0.1\ \mathrm{mHz} \sim  1\ \mathrm{Hz}$\\
Position resolution   &   $\leq 1.8\ \mathrm{nm}\ \mathrm{Hz}^{-1/2}$ \\
Voltage noise  &   $\leq 2.2\times 10^{-14}\ \mathrm{m}\ \mathrm{s}^{-2}\ \mathrm{Hz}^{-1/2}$\\
\end{tabular}
\end{ruledtabular}
\end{table}
The structural configuration of the Taiji-2 SC has not yet been finalized, there are two potential SC c.m. placements. 
In the first configuration, shown in Fig.~\ref{fig:Pos_com1}, SC's c.m. is at the midpoint between the two TMs, the red dot indicates the SC's c.m. location.
In the second configuration, shown in Fig.~\ref{fig:pos_SC_com1}, its c.m. located at the intersection of the two laser links.

The primary scientific data transmitted from the SC to Earth include raw voltage readouts from the electrodes, as shown in Fig.~\ref{fig:Taiji_XYZ}. These voltages, through calibrated scale factors, can be converted into linear and angular accelerations of the TM relative to the electrode cage. 
The corresponding conversion relationships are detailed in Tab.~\ref{tab:com:V_S}. 
Here, $\mathrm{m}$ denotes the mass of the TM and $\mathrm{J}_\phi$, $\mathrm{J}_\theta$ and $\mathrm{J}_\psi$ represents the moment of inertia with respect to the corresponding axes; 
$\mathrm{L}_x$, $\mathrm{L}_y$ and $\mathrm{L}_z$ are the distances from the TM to the electrode plates along respective axis; $\varepsilon_0$ is the vacuum permittivity; $\mathrm{S}_x$, $\mathrm{S}_y$ and $\mathrm{S}_z$ are the areas of the electrode plates associated with each axis; $\mathrm{V}_P$ is the preload bias voltage; $\mathrm{l}_{\phi}$, $\mathrm{l}_{\theta}$ and $\mathrm{l}_{\psi}$ represent the lever arms of the electrostatic force acting on the rotational degrees of freedom for each axis of TM.

\begin{table}[htb]
\renewcommand\arraystretch{2}
\centering
\caption{Acceleration expression and scale factors for each degree of freedom in GRS. All expressions in the table are given in the corresponding TMRF coordinate system. 
Here, $V_{x1}$ and $V_{x2}$ denote the voltage readouts of the two electrode channels for the $x$-axis degree of freedom. 
Similarly, $V_{y1}$ and $V_{y2}$, and $V_{z1}$ and $V_{z2}$, denote the corresponding readouts for the $y$- and $z$-axis degrees of freedom, respectively.}
\label{tab:com:V_S}
    \begin{ruledtabular}
\begin{tabular}{c c c}
	 Axis  & Scale factor $\mathrm{G}_{P}$ &  Acceleration expression\\
	\hline 
 $x$ & $\frac{ \varepsilon_0 \mathrm{S}_x \mathrm{V}_P}{\mathrm{m} \mathrm{L}_x^2}$ & $\mathrm{G}_{P x} \cdot \left(V_{x 1}+V_{x 2}\right)$ \\
	 $y$ & $\frac{ \varepsilon_0 \mathrm{S}_y \mathrm{V}_P}{2 \mathrm{m} \mathrm{L}_y^2}$ & $\mathrm{G}_{P y} \cdot \left(V_{y 1}+V_{y 2}\right)$ \\
	 $z$ & $\frac{ \varepsilon_0 \mathrm{S}_z \mathrm{V}_P}{2 \mathrm{m} \mathrm{L}_z^2}$ & $\mathrm{G}_{P z} \cdot \left(V_{z 1}+V_{z 2}\right)$ \\
	$\phi$        & $\frac{\varepsilon_0 \mathrm{S}_y \mathrm{l}_{\phi} \mathrm{V}_P}{2 \mathrm{J}_{\phi} \mathrm{d}_y^2}$ & $\mathrm{G}_{P \phi} \cdot\left(V_{y 1}-V_{y 2}\right)$ \\
	$\theta$      & $\frac{\varepsilon_0 \mathrm{S}_z \mathrm{l}_\theta \mathrm{V}_P}{2 \mathrm{J}_\theta \mathrm{d}_z^2}$ & $\mathrm{G}_{P \theta} \cdot\left(V_{z 1}-V_{z 2}\right)$ \\
	$\psi$        & $\frac{ \varepsilon_0 \mathrm{S}_x \mathrm{l}_\psi \mathrm{V}_P}{\mathrm{J}_\psi \mathrm{d}_x^2}$ & $\mathrm{G}_{P \psi} \cdot\left(V_{x 1}-V_{x 2}\right)$ \\
\end{tabular}
    \end{ruledtabular}
\end{table}

In the drag-free operational mode, the GRS readouts are transmitted to the control system to compute relative displacement between the TM and the SC along the sensitive axis for compensating non-conservative forces acting on the SC. Precise measurements are required. As shown in Tab.~\ref{tab:com:V_S}, achieving accurate measurements necessitates high-precision scale factors to ensure precise compensation, such that the compensated residual acceleration noise meets the accuracy requirements of Taiji-2.
The precision requirements for the scale factors can be determined based on the residual acceleration noise of the TM and the accuracy requirements of Taiji-2.

Solar radiation pressure (SRP) is the primary non-conservative force acting on the Taiji-2 SC. 
The SC is equipped with dedicated thrusters to compensate for the effects of SRP, but the fluctuations remained.
The fluctuations introduce an acceleration disturbance $S_{\mathrm{\delta SRP}} \leq 1.1\times 10^{-11} (\frac{3\ \mathrm{mHz}}{f}) \mathrm{\ m\ s^{-2}\ Hz^{-1/2}}$ into the GRS~\cite{8-20241376}.
As shown in Tab.~\ref{tab:com:V_S}, the linear acceleration read by the GRS is linearly related to its voltage output, with the slope defined as the scale factors. 
Consequently, voltage fluctuations equivalent to SRP fluctuations denote 
\begin{equation}
S_{\textrm{V}}=\frac{S_{\mathrm{\delta SRP}}}{\mathrm{G_{real}}}, 
\label{eq:S_V}
\end{equation}
where $\mathrm{G_{real}}$ is the true scale factors of the GRS.
When a scale factor calibration error $\Delta \mathrm{G}$ is present, the GRS measurements include an additional acceleration error induced by the coupling between the calibration error and the measured voltage signal:
\begin{equation}
    S_{\Delta \mathrm{a}}=\Delta \mathrm{G} * S_{\textrm{V}}.
    \label{eq:S_a}
\end{equation}

To meet the Taiji-2 precision requirements for the residual acceleration noise of the TM after compensated disturbances via the drag-free control system ($S_{\Delta \mathrm{a}} \leq 3\times10^{-14} \mathrm{\ m\ s^{-2}\ Hz^{-1/2}}$), combined Eq.~\ref{eq:S_V} and Eq.~\ref{eq:S_a}, the calibration error of the scale factors must satisfy the following equation:
\begin{equation}
\frac{\Delta \mathrm{G}}{\mathrm{G_{real}}} = \frac{S_{\Delta \mathrm{a}}}{S_{\mathrm{\delta SRP}}}.
\end{equation}

Putting in corresponding number into the equation reveals that the precision of the estimation of the GRS scale factors must within 0.3\%.

The dynamics of SC used for gravitational wave detection have been thoroughly explored in previous studies~\cite{bortoluzzi2003lisa}. 
A dynamic model is developed, based on previous work and incorporating relevant parameters of Taiji-2, to analyze the impact of scale factors and c.m. offsets variations on the in-orbit performance accuracy of the GRS.
Since this study focuses on in-orbit parameter calibration, given the relatively short duration of the experiment (a few hours), time-varying effects such as solar gravitational drift can be considered constant, and the telescope breathing angle~\cite{JOFFRE20213868} is assumed to be fixed at zero. 
Therefore, the reference frames introduced in the dynamic model are primarily the constellation reference frame and the test mass reference frame. 
The references below are defined and illustrated in Fig.~\ref{fig:Pos_com1} and Fig.~\ref{fig:pos_SC_com1}:
\begin{itemize}

\item Constellation Reference Frame (CRF): This is a reference frame reconstructed on the SC, enabling the SC to determine its attitude relative to the constellation formation. The origin is located at the c.m. of the SC. The axis $\mathrm{Z}$ is defined to be perpendicular to the plane formed by two non-parallel incident laser beams, while $\mathrm{X}$ lies within the same plane and bisects the angle between the two beams. The CRF is defined by the set $\left\{ \mathrm{O}_\mathrm{C}, \mathrm{X}, \mathrm{Y}, \mathrm{Z} \right\}$. 

\item Two Test Mass Reference Frames (TMRF): The origin of the coordinate frame is defined at the nominal center of the GRS electrode cage, three-axis directions are shown in Fig.~\ref{fig:Pos_com1}. The TMRF is defined by the set $\left\{\mathrm{O}_{\mathrm{m}_i}, \mathrm{x}_{i}, \mathrm{y}_{i}, \mathrm{z}_{i}\right\}$, $i =\left\{1,2\right\}$. 
\end{itemize}

The complete expression describes the acceleration of the i-th TM during flight, as observed in the CRF when the system operates in accelerometer mode: 
\begin{equation}
\ddot{\vec{r}}_{\mathrm{m}_i}=\mathrm{m}^{-1}\bigg(\vec{F}_{\mathrm{E}_i}+\vec{D}_{\mathrm{m}_i}+\vec{F}_{\mathrm{S}_i}\bigg),
\label{eq:a_if}
\end{equation}
here $\textrm{m}$ denotes the mass of the TM, $\vec{F}_{\mathrm{E}}$ is electrostatic suspension force, and $\vec{D}_{\mathrm{m}}$ represents the disturbance force acting on TM, the spring-like force $\vec{F}_{\mathrm{S}}$ reads:
\begin{equation}
\vec{F}_{\mathrm{S}_i}= K_{\mathrm{m}_i}{\vec{\delta r}}_{\mathrm{m}_i}, 
\label{eq:f_para}
\end{equation}
each TM is coupled to the SC through a spring-like coupling with a parastic stiffness $K_{\mathrm{m}}$, ${\vec{\delta r}}_{\mathrm{m}}$ is the offsets vector, in the CRF, from the center of the GRS electrode cage to the c.m. of the TM. Under the accelerometer operating mode, the TM is assumed to remain stationary relative to the electrode cage, and thus this term can be neglected. 

Similarly, the acceleration of the SC in the CRF can be expressed as:
\begin{equation}
\ddot{\vec{r}}_{\mathrm{SC}}= {\mathrm{M}}^{-1}(\vec{F}_{\mathrm{T}} + \vec{D}_{\mathrm{SC}}), 
\label{eq:SC_acc}
\end{equation}
where M denotes the mass of SC, $\vec{F}_\mathrm{T}$ is the resultant thruster force applied on SC, and $\vec{D}_\mathrm{S}$ is the disturbance force act on SC.

The attitude expressed in the CRF can be transformed into the TMRF using the following rotation matrix:
\begin{equation}
\begin{aligned}
&\mathrm{C}_{\mathrm{m_1,C}}  = 
&\begin{bmatrix}
\mathrm{cos\alpha/2} & -\mathrm{sin\alpha/2} & 0 \\
\mathrm{sin\alpha/2} & \mathrm{cos\alpha/2} & 0 \\
0 & 0 & 1
\end{bmatrix} 
\end{aligned}
\label{eq:c_s1}
\end{equation}
and
\begin{equation}
\begin{aligned}
&\mathrm{C}_{\mathrm{m_2,C}}  = 
&\begin{bmatrix}
\mathrm{cos\alpha/2} & \mathrm{sin\alpha/2} & 0 \\
-\mathrm{sin\alpha/2} & \mathrm{cos\alpha/2} & 0 \\
0 & 0 & 1
\end{bmatrix}, 
\end{aligned}
\label{eq:c_s2}
\end{equation}
where \(\alpha=\pi/3\) represents the angle between the line-of-sight direction of the telescopes on the two optical platforms in the same SC.

The displacement of TM relative to the electrode housing of the GRS in $\mathrm{TMRF}_i$ could be derived as:
\begin{equation}
\vec{r}_{i}= \mathrm{C}_{\mathrm{m}_i,\textrm{C}} (\vec{r}_{\mathrm{m}_i}-\vec{r}_{\mathrm{SC}}),
\label{eq:r_oi}
\end{equation}
let $\vec{d}_i = \vec{r}_{\mathrm{m}_i}-\vec{r}_{\mathrm{SC}}$, where $\vec{d}_i$ represents the designed offsets between the c.m. of the i-th TM and that of the SC under ideal conditions,
then $\ddot{\vec{r}}_i$ is derived as
\begin{equation}
\begin{aligned}
\ddot{\vec{r}}_i&= \mathrm{C}_{\mathrm{m}_i,\textrm{C}} \ddot{\vec{d}}_i + 2\mathrm{C}_{\mathrm{m}_i,\textrm{C}}\omega_{\mathrm{SC}}\times \dot{\vec{d}}_i + \mathrm{C}_{\mathrm{m}_i,\textrm{C}} \Omega({\omega}_{\mathrm{SC}})\vec{d}_i,
\label{eq:a_mioi}
\end{aligned}
\end{equation}
where \(\Omega({\omega})=\omega_{\mathrm{SC}}\times\omega_{\mathrm{SC}}+\dot{\omega}_{\mathrm{SC}}\), $\omega_{\mathrm{SC}}$ represents angular velocity matrix of SC in the CRF.
Thus, based on Eq.~\ref{eq:a_mioi}, it can be observed that the coupled acceleration noise introduced by the attitude of SC can be write as:
\begin{equation}
{\vec{a}}_{\mathrm{com}_i} \triangleq   2\mathrm{C}_{\mathrm{m}_i,\textrm{C}}\omega_{\mathrm{SC}}\times \dot{\vec{d}}_i + \mathrm{C}_{\mathrm{m}_i,\textrm{C}} \Omega({\omega}_{\mathrm{SC}})\vec{d}_i.
\label{eq:a_com_first}
\end{equation}
Since the variation of the c.m. is extremely slow, $\dot{\vec{d}}_i$ can be neglected compared to $\vec{d}_i$. As a result, the attitude-coupled noise in the GRS readouts is primarily dominated by the second term in Eq.~\ref{eq:a_com_first}.
Thus acceleration noise induced by attitude of SC could be simplified as:
\begin{equation}
\begin{aligned}
{\vec{a}}_{\mathrm{com}_i} =&  \mathrm{C}_{\mathrm{m}_i,\textrm{C}}(\dot{\omega}_{\mathrm{SC}} \times \vec{d}_i+\omega_{\mathrm{SC}} \times(\omega_{\mathrm{SC}} \times \vec{d}_i)). 
\label{eq:a_com_simp}
\end{aligned}
\end{equation}

From Eq.~\ref{eq:a_com_simp}, it can be seen that by carefully calibrating $ \vec{d}_i $, the $\vec{a}_{\mathrm{com}_i} $ signal can be accurately identified and eliminated from the GRS readouts. 
In Eq.~\ref{eq:a_com_simp}, the second term is a second-order small quantity, and thus the attitude couple with the c.m. offsets and generate noise in the linear acceleration primarily dominated by the angular acceleration. Based on the Taiji-2 requirements for residual acceleration noise along the sensitive axis, we can derive the precision requirements for the c.m. bias. 
Specifically, in scientific detection mode, Taiji-2 requires pointing jitter, the residual angular fluctuation of the telescope line-of-sight about its nominal pointing direction, to be below 30~$\mathrm{nrad}\ \mathrm{Hz}^{-1/2}$, implying an angular acceleration jitter of 0.1~$\mathrm{nrad}\ \mathrm{s}^{-2}\ \mathrm{Hz}^{-1/2}\ @\ 10\ \mathrm{mHz}$, under this pointing jitter, the residual acceleration introduced by the c.m. offsets residual is given by:
\begin{equation}
\vec{a}_{\Delta \vec{d}} \approx \dot{\omega}_{\mathrm{jit}} \times \Delta \vec{d},
\end{equation}
$\Delta \vec{d}$ represents the residual error between the true c.m. value and the calibrated value, $ \dot{\omega}_{\mathrm{jit}}$ is the angular acceleration introduced by pointing jitter. Based on the requirements of Taiji-2, the in-orbit c.m. calibration accuracy must reach within 100 $\upmu\textrm{m}$.

\section{calibration method for GRS in accelerometer mode}
\label{sec:offset_scheme}

\begin{figure}[htb]
\centering
\includegraphics[width=.23\textwidth]{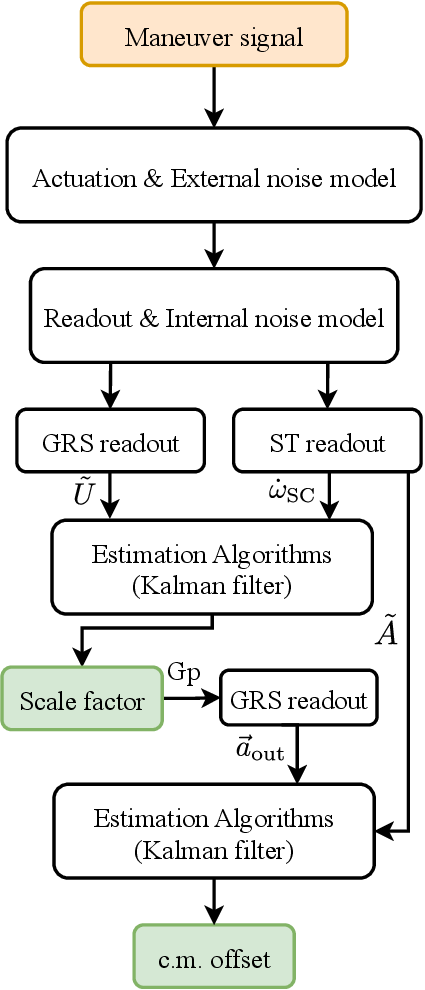}
\caption{Flowchart of Taiji-2 calibration scheme.}
\label{fig:cali alg}
\end{figure}

Taiji-1 proposed a scheme utilizing SC maneuvers in accelerometer mode to calibrate scale factors and c.m. offsets~\cite{zhang2023systematic,PhysRevD.108.082001}.
Taiji-2 adopts a similar in-orbit calibration procedure, as illustrated in Fig.~\ref{fig:cali alg}, the SC swings maneuvers by applying a square wave signal through thrusters, the maneuver signal recorded by both ST and GRS. 
To save propellant consumption, this study uses a single experiment to simultaneously calibrate the scale factors and c.m. offsets.
The scale factors can be calibrated by correlating angular acceleration measurements from ST with voltage signal readouts from the GRS.
The calibrated scale factors are incorporated into the GRS readouts, which are then combined with the ST readouts to calibrate the c.m. offsets.
To clarify the relation between the two calibrations, we note that the c.m. offset estimation is performed using the physical acceleration reconstructed from the GRS voltage readouts. 
Specifically, we first calibrate the GRS scale factors and then apply these calibrated scale factors to convert the corresponding voltage combinations into linear accelerations. 
Therefore, the uncertainty in the scale factor estimation propagates into the c.m. offset calibration through this acceleration conversion step. 

\begin{figure}[ht]
\includegraphics[width=.5\textwidth]{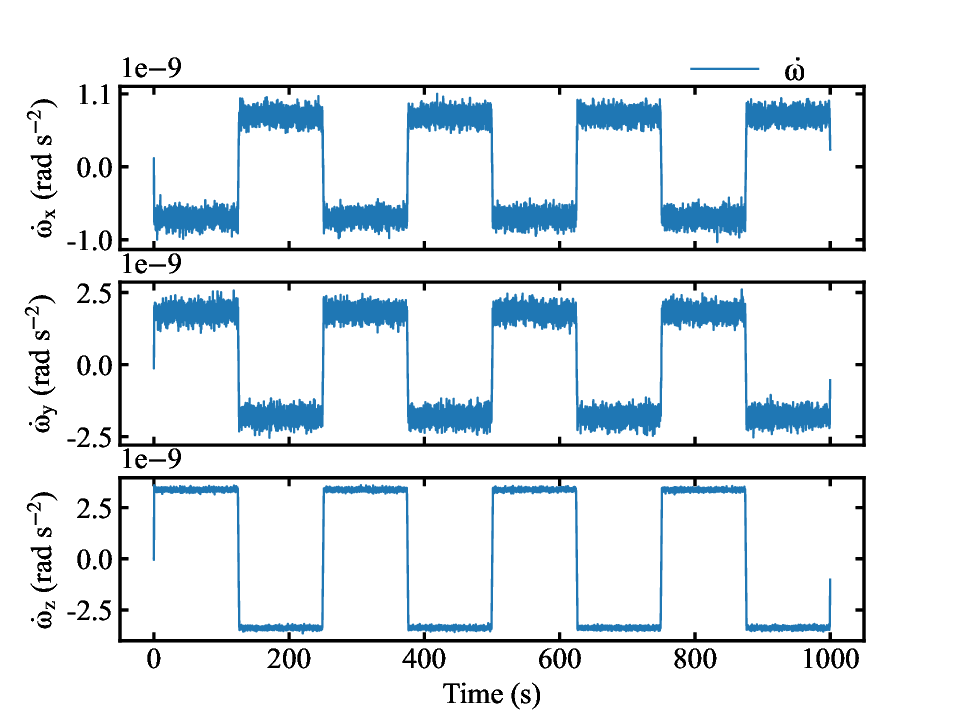}
\caption{The simulated ST readout of maneuvering angular acceleration square wave for Type-1 SC layout. Only a 1000s segment is shown.}
\label{fig:dot_omega_type1}
\end{figure}

\begin{figure}[ht]
\centering
\includegraphics[width=.5\textwidth]{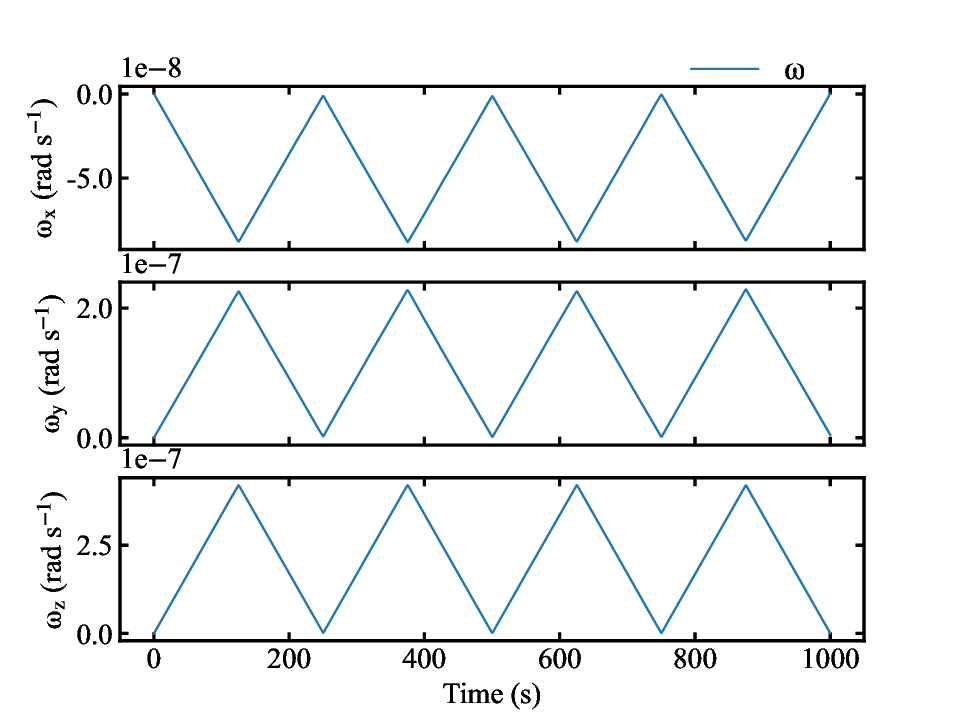}
\caption{The simulated ST readout of maneuvering angular velocity signal for Type-1 SC layout. Only a 1000s segment is shown.}
\label{fig:omega_type1}
\end{figure}

For the in-orbit calibration, the first step following Fig.~\ref{fig:cali alg} is injecting SC attitude oscillation.
The resulting attitude motion is measured by the ST. 
Using the Type-1 spacecraft layout and the corresponding calibration signal, and taking into account the ST readout noise, we simulated the SC angular velocity and angular acceleration readouts in the CRF, as shown in Fig.~\ref{fig:dot_omega_type1} and Fig.~\ref{fig:omega_type1}. 
For clarity, only a 1000-second segment is displayed here.
Type-2 has similar results and is not shown here.

The next step is to estimate the scale factors using the ST readout of the SC angular acceleration $\dot{\omega}_{\mathrm{SC}}$ together with the GRS voltage readouts $\tilde{U}$, using a Kalman filter as indicated in Fig.~\ref{fig:cali alg}. 
The angular acceleration of SC in CRF calculated from the GRS readouts can be expressed as
\begin{equation}
\dot{\omega}_\textrm{SC}=\mathrm{C}_{\mathrm{m}_i,\mathrm{C}}^{-1}\tilde{U} \mathrm{G}_{\mathrm{angle}},
\label{eq:dot_omega_out}
\end{equation}
here, $\mathrm{G}_{\mathrm{angle}}=[\mathrm{G}_{P\phi},\mathrm{G}_{P\theta},\mathrm{G}_{P\psi}]^{\mathrm{T}}$, and $\tilde{U}$ can be expressed as
\begin{equation}
\tilde{U}=\left[\begin{array}{ccc}
V_{Y 1}-V_{Y 2} & 0 & 0 \\
0 & V_{Z 1}-V_{Z 2} & 0 \\
0 & 0 & V_{X 1}-V_{X 2}
\end{array}\right].
\label{eq:U_tilde}
\end{equation}
Combined with the ST readouts, the scale factors $\mathrm{G}_{\mathrm{angle}}$ can be estimated, this paper uses the CRF coordinate system for scale factors estimation.

As shown in Fig.~\ref{fig:cali alg}, the calibrated scale factors are then applied to reconstruct the GRS linear acceleration $\vec{a}_{\mathrm{out}}$. 
The ST readouts are further used to form the angular velocity and angular acceleration mapping matrix $\tilde{A}$. 
Using $\vec{a}_{\mathrm{out}}$ together with $\tilde{A}$, the c.m. offset is subsequently estimated via Kalman filtering.
The c.m. offsets calibration is performed in the TMRF.
The accelerometer measurement model can be expressed as follows,
\begin{equation}
\vec{a}_{\textrm{out}_i}=\mathrm{C}_{\mathrm{m}_i,\textrm{C}}( \tilde{A} \vec{d}_{i}+\Gamma t+\Lambda)+A_\textrm{n},
\label{eq:A_out}
\end{equation}
where $\tilde{A}$ can be expressed as
\begin{equation}
\tilde{A}=\left[\begin{array}{ccc}
-\omega_\textrm{y}^{2}-\omega_\textrm{z}^{2} & \omega_\textrm{x} \omega_\textrm{y}-\dot{\omega}_\textrm{z} & \omega_\textrm{x} \omega_\textrm{z}+\dot{\omega}_\textrm{y} \\
\omega_\textrm{x} \omega_\textrm{y}+\dot{\omega}_\textrm{z} & -\omega_\textrm{x}^{2}-\omega_\textrm{z}^{2} & \omega_\textrm{z} \omega_\textrm{y}-\dot{\omega}_\textrm{x} \\
\omega_\textrm{x} \omega_\textrm{z}-\dot{\omega}_\textrm{y} & \omega_\textrm{z} \omega_\textrm{y}+\dot{\omega}_\textrm{x} & -\omega_\textrm{y}^{2}-\omega_\textrm{x}^{2}
\end{array}\right].
\label{eq:A_tilde}
\end{equation}
The corresponding axis of the SC can be represented by $\omega_\textrm{i}$ and $\dot{\omega}_\textrm{i}$ ($\textrm{i}=\textrm{x},\textrm{y},\textrm{z}$) for angular acceleration and angular velocity, respectively. 
The linear slope is represented by $\Gamma$ and the constant bias by $\Lambda$. 
In this study, we removed the linear effect by detrending $\vec{a}_\textrm{out}$. Therefore, the model of $\vec{a}_\textrm{out}$ utilized in this study is,
\begin{equation}
\vec{a}_{\textrm{out}_i}=\mathrm{C}_{\mathrm{m}_i,\textrm{C}} \tilde{A} \vec{d}_i + A_\textrm{n}, 
\label{eq:used_A_out}
\end{equation}
where $A_\textrm{n}$ is the measurement noise.
Combined with the angular velocity and angular acceleration of the SC, the c.m. offsets can be estimated.

This paper employs the Kalman filter~\cite{10.1115/1.3662552,10.1115/1.3425006,bucy1971digital} to estimate the parameters of GRS.
The model for calculating the SC’s angular acceleration from the GRS readouts is provided in Eq.~\ref{eq:dot_omega_out}; 
the accelerometer measurement model is given by Eq.~\ref{eq:used_A_out}. 
The unknown parameters are treated as the state vector $X$. 
Here, $X = [\mathrm{G}_{P\phi},\mathrm{G}_{P\theta},\mathrm{G}_{P\psi}]^{\mathrm T}$ for scale factors calibration; 
for the c.m. offsets calibration, $X = [d_\textrm{x}, d_\textrm{y}, d_\textrm{z}]^{\mathrm T}$ with $\dot{d_\textrm{i}} = 0$ for $\textrm{i}=\textrm{x},\textrm{y},\textrm{z}$. 
The state equation, as derived in Ref.~\cite{Dong2009}, is,
\begin{equation}
\hat{X}_{k}=\Phi_{k,k-1} X_{k-1}, \qquad \Phi_{k,k-1}=I.
\end{equation}
Here, $\hat{X}$ represents the prior value, $k$ signify the step of filter, $\Phi_{k,k-1}$ represents the state transition matrix from step $k-1$ to step $k$,  
and $I$ is the identity matrix as no manipulation is performed on the state vector.

The output of the GRS can be defined as the observation equation,
\begin{equation}
Z_\textrm{out,\textit{k}}=H_{k} {X}_{k}+V_{k},
\end{equation}
in the scale factors calibration, $H_{k}=\tilde{U}_{k}$; in the c.m. offsets calibration, $H_{k}=\tilde{A}_{k}$, and $V_{k}$ is the discrete measurement noise that satisfies,
\begin{equation}
E\left\{V_{k}\right\}=0, Cov\left\{V_{k}\right\}=R_{k}.
\end{equation}
Here, $R_{k}$ denotes the variance matrix of the measurement noise,
assuming that all $V_{k}$ are independent, unbiased, and possess finite variance which implies that $R_{k}$ is a diagonal matrix, the value of $R_{k}$ also varies in different calibrations.

The Kalman filter recursion is summarized as: 
\begin{equation}
\hat{Z}_\textrm{out,\textit{k}}=H_{k} \hat{X}_{k},
\end{equation}
\begin{equation}
\hat{P}_{k}=\Phi_{k,k-1} P_{k-1} \Phi_{k,k-1}^{\mathrm{T}}+Q_{k-1},
\end{equation}
\begin{equation}
K_{k}=\hat{P}_{k} H_{k}^{\mathrm{T}}\left(H_{k} \hat{P}_{k} H_{k}^{\textrm{T}}+R_{k}\right)^\mathrm{-1},
\end{equation}
\begin{equation}
X_{k}=\hat{X}_{k}+K_{k}\left(Z_\textrm{out,\textit{k}} - \hat{Z}_\textrm{out,\textit{k}}\right),
\end{equation}
\begin{equation}
P_{k}=\left(I-K_{k} H_{k}\right) \hat{P}_{k}\left(I-K_{k} H_{k}\right)^{\mathrm{T}}+K_{k} R_{k} K_{k}^{\mathrm{T}},
\end{equation}
where $\hat{Z}_\textrm{out,\textit{k}}$ is the predicted observation, $\hat{P}_{k}$ is the estimated error covariance, the process noise covariance is set to $Q_{k-1}=0$ in this study, $K_k$ is the Kalman gain, $X_k$ is the posterior estimate, and $P_k$ is the posterior error covariance. 
After filtering, the optimal estimates and their uncertainties are obtained from $X_k$ and $P_k$.

The noise components in the calibration experiment include payload noise, SC platform noise and space environmental noise. 
The payload noise primarily consists of readout noise from the GRS and ST, as well as noise from the microthrusters and the residual noise of low frequency suspension (LFS) control system in accelerometer mode. 
Table.~\ref{tab:payload_n} details the payload noise considered during the calibration experiment, LFS translational ($\mathrm{LFST}$) and LFS rotational ($\mathrm{LFSR}$) respectively denote the residual translational and rotational acceleration noises from LFS control.
The total environmental noise of the SC platform that affects the TMs, including residual gases, temperature gradients, stray electric fields, and residual charges, etc., is set to $3\times10^{-14}\ \mathrm{m}\ \mathrm{s}^{-2}\ \mathrm{Hz}^{-1/2}$. 
For the noises that act on the SC from the space environment, the primary disturbance is the fluctuation of solar radiation pressure, with a acceleration noise level about  $1.1\times 10^{-11} (\frac{3\ \mathrm{mHz}}{f}) \mathrm{\ m\ s^{-2}\ Hz^{-1/2}}$.

\begin{table}[htb]
\begin{center}
\caption{The payload noise considered during the calibration experiment. $\mathrm{LFST}$ and $\mathrm{LFSR}$ respectively denote the residual translational and rotational acceleration noises from LFS control.}
\label{tab:payload_n}
\begin{ruledtabular}
\begin{tabular}{cc}
Noise sources &   Nominal Value\\
\hline
ST readout noise  &   $\leq 0.2^{\prime \prime}$\\
Thruster noise   &   $\leq 0.1\ \mathrm{\upmu N}\ \mathrm{Hz}^{-1/2}$\\
$\mathrm{LFST}_x$    &    $\leq 2.4\times 10^{-12}\ \mathrm{m}\ \mathrm{s}^{-2}\ \mathrm{Hz}^{-1/2}$\\
$\mathrm{LFST}_y$,$\mathrm{LFST}_z$      &      $\leq 3.6\times 10^{-12}\ \mathrm{m}\ \mathrm{s}^{-2}\ \mathrm{Hz}^{-1/2}$\\
$\mathrm{LFSR}_x$   &   $\leq 2\times 10^{-11}\ \mathrm{rad}\ \mathrm{s}^{-2}\ \mathrm{Hz}^{-1/2}$\\
$\mathrm{LFSR}_y$,$\mathrm{LFSR}_z$   &   $\leq 3.5\times 10^{-11}\ \mathrm{rad}\ \mathrm{s}^{-2}\ \mathrm{Hz}^{-1/2}$\\
\end{tabular}
\end{ruledtabular}
\end{center}
\end{table}

Based on the previous discussions, a parameterized calibration scheme, estimation algorithm, and simulation approach can be formulated considering dynamic characteristics and various noise models of Taiji-2.
The calibration experiment of Taiji-2 employs thrusters to maneuver the SC.

The c.m. offsets between the TMs and the SC are denoted by $\vec{d}_i$ in the TMRF. 
Using Eq.~\ref{eq:a_com_simp}, the linear acceleration induced by the coupling between the optimized maneuver and the c.m. offsets, as observed by the GRS, can be estimated~\cite{wang2010determination,rs14164030}.

The line acceleration introduced by the calibration experiment combined with the background noise must falls within the measurable range of the GRS and the thrust should align with the capabilities of micro-thrusters (0.1-100~$\upmu \mathrm{N}$).
The design performance of the Taiji-2 GRS exhibits a critical point around 3 mHz, where performance declines below this frequency. However, considering that longer maneuver periods can yield larger SC swing angles, thereby enhancing the signal-to-noise ratio of the SC attitude signals, this calibration experiment is designed with a square wave period of 250-seconds.
To meet the signal transmission requirements between the SC and ground station, all simulations in this study use a sampling rate of 10 Hz.

From Eq.~\ref{eq:a_com_simp}, it can be derived that maximizing SC angular acceleration during maneuvers aims to enhance the GRS signal-to-noise ratio.
Due to the differences in the c.m. of TMs distribution structures of the two type SC configurations, and to meet the GRS dynamic control range of $1\times 10^{-9}\ \mathrm{m}\ \mathrm{s^{-2}}$, 
this study designs distinct maneuver schemes based on the different c.m. of TMs configurations. For the Type-1 SC layout, the total thrust amplitudes applied in the three axes are $3\ \upmu \mathrm{N\ m}$, $3\ \upmu \mathrm{N\ m}$ and $13\ \upmu \mathrm{N\ m}$ respectively. 
For the Type-2 SC layout, the thrusts applied in the three axes are $11\ \upmu \mathrm{N\ m}$, $3\ \upmu \mathrm{N\ m}$ and $19\ \upmu \mathrm{N\ m}$.
By combining the rotational inertia matrix $\mathrm{J}$ of SC with the applied thruster noise, the angular acceleration and angular velocity of the SC can be obtained.

\section{calibration results}
\label{sec:result}

\begin{figure}[H]
\centering
\includegraphics[width=.48\textwidth]{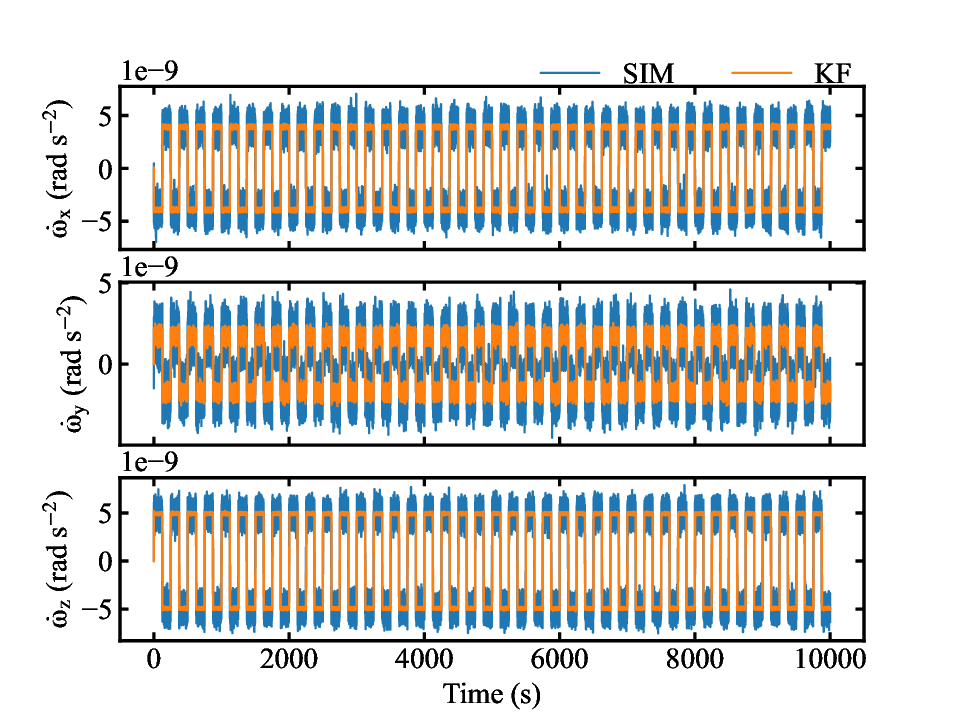}
\caption{The fit results of GRS1 and simulated ST readouts for Type-1.}
\label{fig:KF_alpha_fit_GRS1_type1}
\end{figure}

\begin{figure}[H]
\centering
\includegraphics[width=.48\textwidth]{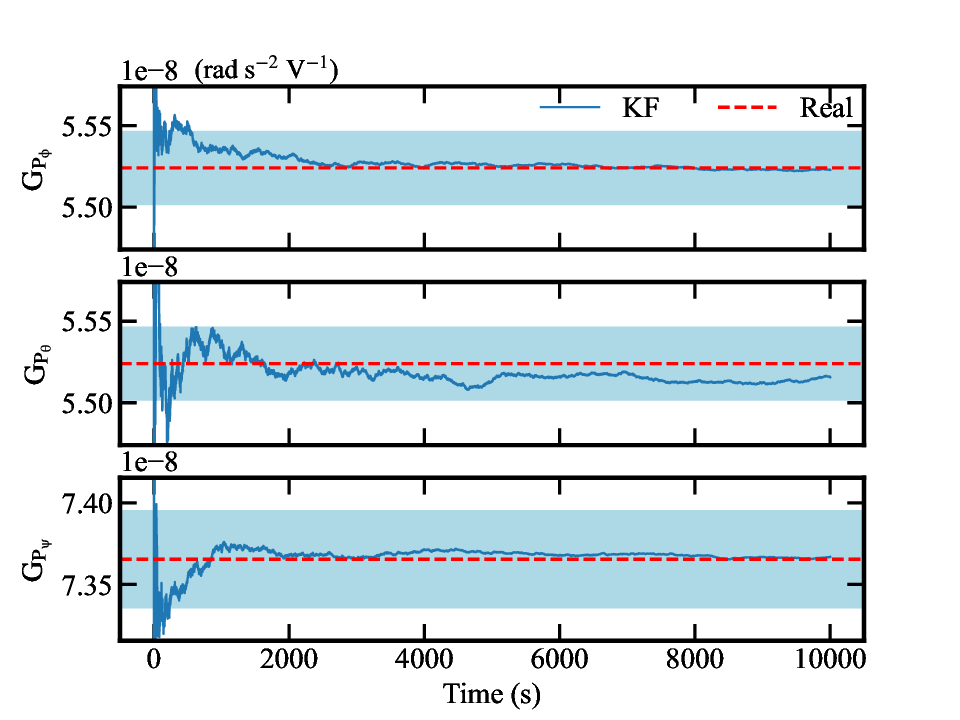}
\caption{The calibration results of scale factors of GRS1 for Type-1.  The blue shaded area indicates the Taiji-2 requirement on the scale factor estimation error ($<0.3\%$).}
\label{fig:KF_alpha_result_GRS1_type1}
\end{figure}

The angular acceleration fitting results for estimating the GRS1 scale factors of the Type-1 SC layout using the Kalman filter are shown in Fig.~\ref{fig:KF_alpha_fit_GRS1_type1}. 
Here, SIM denotes the simulated readouts, and KF represents the results obtained using the Kalman filter. 
In Fig.~\ref{fig:KF_alpha_fit_GRS1_type1}, the blue line represents the simulation readouts from the ST. 
The orange line indicates the fitted results obtained by combining the GRS readouts with the Kalman filter. 
The Kalman filter results of $\mathrm{G_{angle}}$ are shown in Fig.~\ref{fig:KF_alpha_result_GRS1_type1}, the blue line represents the estimated process of the scale factors. 
The red dashed line indicates the true value. 
The blue shaded area represents the range of scale factor estimation error requirements for Taiji-2 ($\le 0.3\%$).

Since the scale factors results of the two GRS of the same SC layout are largely consistent, only the estimating process and results for GRS1 is presented here. 
The estimation processes for the Type-2 SC layout are not displayed.
The detailed final results can be found in Tab.~\ref{tab:result_scale3}, ES represents estimated results, 
the Taiji-2 requirement on the scale factor estimation error is $\le 0.3\%$, and the corresponding absolute values are given in the second row, 
the upper part of Tab.~\ref{tab:result_scale3} shows the estimation results for Type-1 SC layout, while the lower part presents the results for Type-2 SC layout, the calibration errors of the scale factors along all three axes are below 0.2\%, satisfying the calibration requirements of the Taiji-2 mission.
It confirms a consistent linear conversion between angular and linear scale factors (Tab.~\ref{tab:com:V_S}), ensuring GRS's linear acceleration scale factors also meet Taiji-2's requirements.

\begin{table}[htbp!]
    \footnotesize
	\caption{The estimation results of scale factors of GRS. The Taiji-2 requirement on the scale factor estimation error is $\le 0.3\%$, and the corresponding absolute values are given in the second row.}
        \begin{ruledtabular}
			\begin{tabular}{cccc}
                Scale Factor & $\mathrm{G}_{P \phi} (\times 10^{-8})$ & $\mathrm{G}_{P \theta}(\times 10^{-8})$ & $\mathrm{G}_{P \psi}(\times 10^{-8})$ \\
                \hline
                Error Requirement & $\leq$0.017 & $\leq$0.017 & $\leq$0.022 \\
				\hline
			    Truth (Type-1)    & 5.524   & 5.524   &  7.365\\
                ES (GRS1)       & 5.523$\pm$0.004 & 5.516$\pm$0.010 &  7.367$\pm$0.005\\
                ES (GRS2)       & 5.523$\pm$0.004 & 5.516$\pm$0.010 & 7.366$\pm$0.005\\
                \hline
			    Truth (Type-2)    & 5.524   & 5.524   &  7.365\\
                ES (GRS1)       & 5.523$\pm$0.002 & 5.525$\pm$0.005 &  7.363$\pm$0.002\\
                ES (GRS2)       & 5.523$\pm$0.002 & 5.525$\pm$0.005 & 7.363$\pm$0.002\\
	\end{tabular}
        \end{ruledtabular}
	\label{tab:result_scale3}
\end{table}

\begin{figure}[htbp!]
\centering
\includegraphics[width=.48\textwidth]{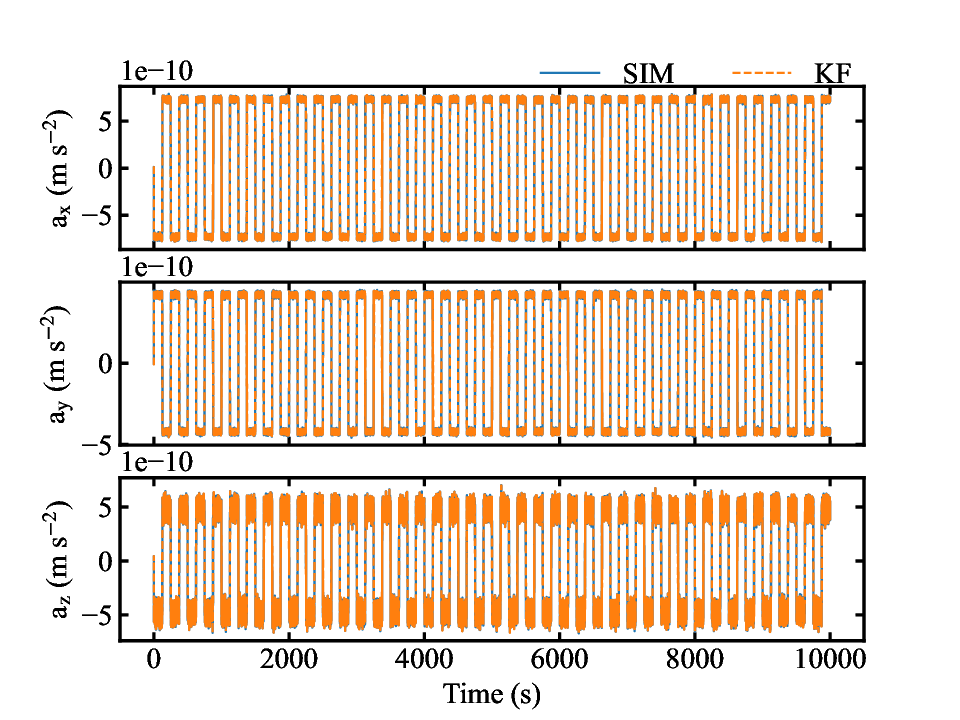}
\caption{The fit results of simulated GRS1 readouts for Type-1.}
\label{fig:KF_GRS1_out_type1}
\end{figure}

\begin{figure}[htbp!]
\centering
\includegraphics[width=.48\textwidth]{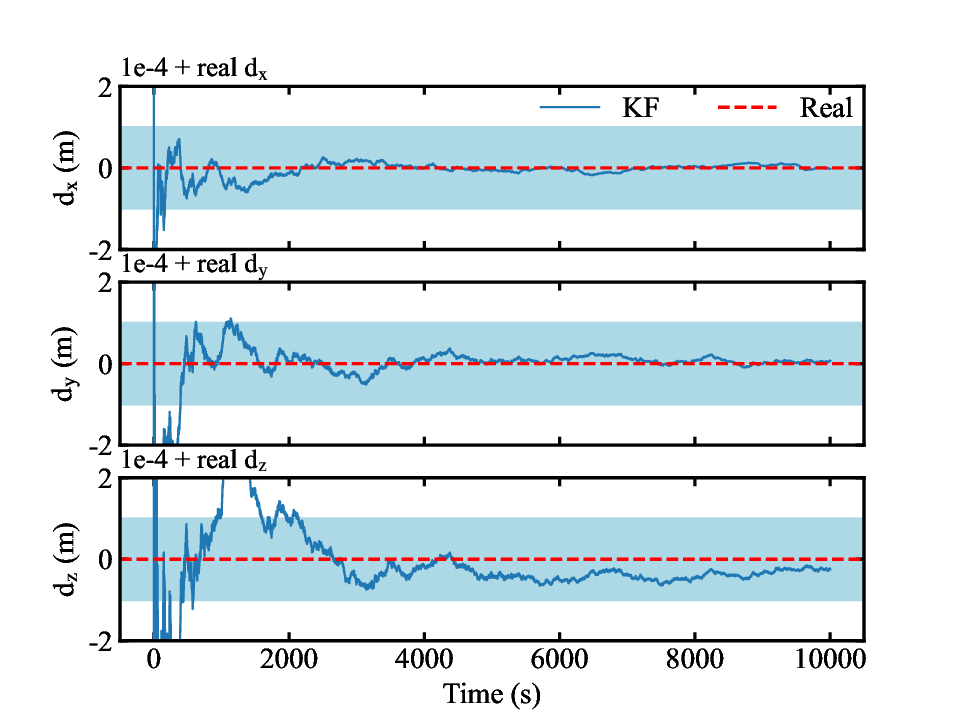}
\caption{The c.m. offsets calibration results of GRS1 for Type-1. The blue shaded area indicates the Taiji-2 requirement on the c.m. offset calibration estimation error ($\le 100~\upmu\mathrm{m}$).}
\label{fig:KF_GRS1_d_type1}
\end{figure}

Figure~\ref{fig:KF_GRS1_out_type1} and Figure~\ref{fig:KF_GRS1_d_type1} display the fitted and estimated results for GRS1 concerning the c.m. offsets between TM and the SC for Type-1  SC layout.
In Fig.~\ref{fig:KF_GRS1_out_type1}, the blue line represents the simulated GRS readouts, the orange line indicates the fitted results through the Kalman filter.
In Fig.~\ref{fig:KF_GRS1_d_type1}, the blue line represents the estimated process of the c.m. offsets, while the red dashed line indicates the true value and the blue shaded area represents the Taiji-2 requirement range for the c.m. offset estimation error ($\le 100~\upmu\mathrm{m}$).

Since the c.m. offsets calibration results of the two GRS of the same SC layout type are largely consistent, only the fitting process and results for GRS1 is presented here. 
The detailed final results can be found in Tab.~\ref{tab:result}, the second row lists the Taiji-2 requirement on the c.m. offset calibration estimation error.
The upper part of Tab.~\ref{tab:result} shows the estimation results for Type-1 SC layout, while the lower part presents the results for Type-2 SC layout.
The estimated errors for c.m. offsets along all three axes are within 100$\ \upmu \mathrm{m}$, which meets the requirements of Taiji-2.

\begin{table}[H]
    \footnotesize
    \caption{The results of estimation of the c.m. offsets. The Taiji-2 requirement on the c.m. offset estimation error is $\le 100 \upmu$m.}
    \begin{ruledtabular}
			\begin{tabular}{cccc}
				Axis & X ($\upmu$m) & Y ($\upmu$m) & Z ($\upmu$m) \\
                \hline
                Error Requirement & $\leq$ 100 &  $\leq$ 100 &  $\leq$ 100 \\
                \colrule
			    Truth (GRS1)    & 100   & 180000   &  200\\
                ES (GRS1)       & 121.0$\pm$70.3 & 179991.6$\pm$31.2 &  172.3$\pm$68.8\\
			    Truth (GRS2)    & 100   & -180000   &  200\\
                ES (GRS2)       & 126.6$\pm$67.4 & -180008.0$\pm$29.9 &  166.7$\pm$70.5\\
                \hline
                Truth (GRS1)    & 216400   & 125100   &  200\\
                ES (GRS1)       & 216397.7$\pm$21.9 & 125106.7$\pm$53.1 &  175.4$\pm$68.2\\
			    Truth (GRS2)    & 216400   & -125100   &  200\\
                ES (GRS2)       & 216397.7$\pm$ 21.4 & -125093.2$\pm$53.1 &  175.5$\pm$68.2\\
	\end{tabular}
    \end{ruledtabular}
\label{tab:result}
\end{table}

\section{Conclusion}
\label{sec:discussion}
The GRS readouts, initially recorded as voltage readouts, require scale factors to convert them into accurate physical measurements. 
Accurate calibration of the scale factors is essential in scientific model to precisely quantify environmental disturbances and fed into the drag-free control system  for compensation.
Precise calibration of the c.m. offsets is essential for suppressing acceleration noise induced by SC attitude motion, especially in high-precision measurement systems.
These calibration procedures are crucial for enhancing accuracy of high-precision GRSs, particularly in space-based gravitational wave observatories where stringent performance is required to meet scientific objectives.

For two potential SC layouts of Taiji-2, we designed targeted in-orbit calibration schemes and conducted simulation validation analyses for each configuration.
A periodic square-wave signal was applied to the SC via thrusters, and Kalman filter was employed to simultaneously estimate both the GRS scale factors and the c.m. offsets. 
The results showed that the scale factors estimation error is within 0.2\% and  the c.m. offsets estimation error is within 100 $\upmu\mathrm{m}$, satisfying the corresponding Taiji-2 calibration requirements. 

For the future, higher-precision Taiji-3 mission, the quality of measurement data will be even more sensitive to biases in the GRS scale factors and the c.m. offsets between the TMs and SC. 
The accuracy of in-orbit calibration for these operational parameters will be critical to achieving the mission’s scientific objectives. 
The calibration principles and methodologies presented in this paper remain applicable. 

\begin{acknowledgments}
This work is supported by the National Key Research and Development Program of China No. 2020YFC2200601, No. 2020YFC2200603, and No. 2021YFC2201901, and the International
Partnership Program of the Chinese Academy of Sciences, Grant No. 025GJHZ2023106GC.
\end{acknowledgments}

\end{document}